\documentclass[aps,notitlepage, floatfix, prx]{revtex4-1}%
\usepackage{amssymb}
\usepackage{amsfonts}
\usepackage{amsmath}
\usepackage{graphicx}
\usepackage{enumerate}
\usepackage[usenames,dvipsnames]{xcolor}
\providecommand{\U}[1]{\protect\rule{.1in}{.1in}}
\begin{document}
\title{Mechanism for the stabilization of protein clusters above the solubility
curve: the role of non-ideal chemical reactions}
\author{James F. Lutsko}
\email{jlutsko@ulb.ac.be}
\homepage{http://www.lutsko.com}
%\author{Pierre Gaspard}
%\author{Gr\'{e}goire Nicolis}
\affiliation{Center for Nonlinear Phenomena and Complex Systems, Code Postal 231,
Universit\'{e} Libre de Bruxelles, Blvd. du Triomphe, 1050 Brussels, Belgium}

\begin{abstract}
Dense protein clusters are known to play an important role in nucleation of protein crystals from dilute solutions. While these have generally been thought to be formed from a metastable phase, the observation of similar, if not identical, clusters above the critical point for the dilute-solution/strong-solution phase transition has thrown this into doubt. Furthermore, the observed clusters are stable for relatively long times. Because protein aggregation plays an important role in some pathologies, understanding the nature of such clusters is an important problem. One mechanism for the stabilization of such structures was proposed by Pan, Vekilov and Lubchenko and was investigated using a DDFT model which confirmed the viability of the model. Here, we revisit that model and incorporate additional physics in the form of state-dependent reaction rates. We show by a combination of numerical results and general arguments that the state-dependent rates disrupt the stability mechanism. Finally, we argue that the state-depedent reactions  correct unphysical aspects of the model with ideal (state-independent) reactions and that this necessarily leads to the failure of the proposed mechanism.

\end{abstract}
\date{\today}

\maketitle

\section{Introduction}

Determining the structure of proteins remains one of the most important
problems in molecular biology. The most common experimental technique involves
x-ray diffraction which is only possible given the availability of
sufficiently large, high-quality crystals. This was one of the main reasons
that protein crystallization became the subject of intense study in the
physical-chemistry community. Protein crystallization is of also of interest
because proteins are large molecules that move slowly in solution and so their
behavior can be monitored at the molecular scale, thus making it possible to
follow processes like nucleation and crystal growth in unprecedented
detail\cite{MikePNAS, Mike_PRL}. One consequence of the intense
interest in this subject was the discovery of the so-called "two-step"
mechanism of protein nucleation wherein the formation of crystals from a dilute
solution involves nucleation of an intermediate metastable phase of dense, but
disordered, solution. The importance of this non-classical nucleation pathway
was first noted in simulations\cite{tWF} and later observed in experiment\cite{VekilovCGDReview2004}.
Further theoretical work suggested that the mechanism is not specific to
proteins and may be of general applicability\cite{LN}. Since then,
two-stage nucleation has been reported in a wide variety of systems.

It was widely believed that the intermediate clusters in the two-step
mechanism were composed of a metastable dense phase: the transition from the
dilute solution to this dense phase being analogous to the vapor-liquid
transition in a simple fluid. While this may indeed be the case, the Vekilov
group reported the observation of similar - if not identical - clusters
occurring \textit{above} the critical temperature of the
dilute-solution/dense-solution transition\cite{VekilovClusters_Evidence1, VekilovClusters_Evidence2, VekilovClusters_Evidence3}. Since there is
only a single disordered phase at such temperatures, the nature of these
(possibly new type of) clusters was a mystery. Several other observations only
compounded the mystery:\ (a) the protein clusters are characterized by a
typical size: i.e. they do not grow as one would expect if they represented
the formation of a new, thermodynamically favored phase; (b) the purity of the
protein solutions seems to rule out impurity effects such as stabilize water
droplets in the atmosphere\cite{Kohler, Kohler1}; (c) the clusters appear to be liquid
in nature; (d) increasing the protein concentration does not appear to affect
the size of the clusters and (e) if the droplets are filtered out of a
solution, new droplets spontaneous form. Given that protein aggregation plays
a role in various pathologies\cite{VekilovClusters1}, the explanation of these
observations is obviously of considerable interest.

Pan, Lubchenko and Vekilov proposed a possible mechanism for the stabilization of
such clusters\cite{VekilovClusters1,VekilovClusters2}. They first suggest that the proteins form a novel complex, the
nature of which is unknown:\ examples are either a sub-population of
mis-folded proteins or a sub-population of oligomers. Whatever they are, the
idea is that the complex is in equilibrium with the population of (ordinary)
protein monomers due to a chemical reaction whereby proteins can be converted
to complexes and vice-versa. Next, they hypothesize that the complexes have a
different phase diagram and in fact are metastable with respect to
condensation into a dense solution. Normally, this would suggest that once
formed, such droplets of dense-solution complex would grow indefinitely.
However, the third element of the hypothetical mechanism is that the droplets
are stabilized by the protein-complex reaction.

In order to investigate whether this mechanism could lead to stable clusters,
we have constructed a Dynamic Density\ Functional Theory model based on it\cite{LN2}. In
order to be concrete, we imagine that the complexes are simply dimers and that
the monomers and dimers interact only via excluded volume. The monomers were
taken to be hard spheres (since we are only interested in them above the
critical point) and the dimers were assumed to interact via a Lennard-Jones
potential tuned so that a vapor-liquid transition was possible. We found that,
indeed, clusters were stabilized due to the fact that since the clusters are,
by definition, dimer-rich, there is preferential production of monomers within
the clusters. This turns out to become more important as the clusters grow so
that at some point, growth stops and the clusters are stable. The physics was
explained based on a simple capillary model leading to simple relations
between the reaction rates and the cluster size and growth rate. Thus, our
model seems to confirm the viability of the mechanism.

Here, we return to this model system in order to incorporate a missing element
of the physics. In our previous work, we assumed that the reaction rates were
constants, independent of the environment. This is unrealistic on physical
grounds. Imagine a collection of molecules that are to undergo a chemical
reaction to form a product. This is typically viewed as a barrier
crossing problem in which the molecules have some initial free energy
$\phi_{i}$, cross a free energy barrier of height $\phi_{B}$, and eventually
arrive at the products which have free energy $\phi_{p}$. The rate at which
the process occurs in the forward direction should be proportional to
$e^{-\beta\left(  \phi_{B}-\phi_{i}\right)  }$ and the rate for the reverse
reaction proportional to $e^{-\beta\left(  \phi_{B}-\phi_{p}\right)  }$ so
that the ratio of the forward to the backward rates is proportional to
$e^{-\beta\left(  \phi_{f}-\phi_{i}\right)  }$, thus giving detailed balance.
This suggests, for example, that relative to the rates in at low densities,
the dimer to monomer reaction should be suppressed in the condensed dimer
phase since this is, by hypothesis, a low-free energy phase compared to the
dilute-dimer solution. Even more telling is that if, as we hypothesized, the
monomers see the dimers as being hard spheres, a monomer in the dimer droplet
will see itself as being a hard-sphere in a dense hard-sphere gas, which means
it will entail a large free energy cost. Hence the monomer to dimer transition
should be enhanced within the droplet. Both of these facts threaten the
proposed stability mechanism.

In the next Section, we describe the basic elements of our original model and
summarize the physical results previously reported. In Section III, we then
generalize the model to include state-dependent reaction rates. We then report
the results of numerical solution of the resulting generalized DDFT model and
interpret them in the context of the capillary model. We conclude with a
discussion of our results and their interpretation in terms of general
principles of nonequilibrium thermodynamics. Our main conclusion is that
state-dependent reactions make the proposed mechanism untenable.

\section{DDFT model with fixed reaction rates}

To be concrete, we take the protein complex to be dimers. If the density of
monomers is $n_{1}$ and that of dimers is $n_{2}$ then we take the reaction
kinetics to be given by
\begin{align}
\frac{dn_{1}}{dt} &  =-2k_{1}n_{1}^{2}+2k_{2}n_{2}\\
\frac{dn_{2}}{dt} &  =k_{1}n_{1}^{2}-k_{2}n_{2}\nonumber
\end{align}
where the stochiometric factors take account of the fact that production of a
dimer requires two monomers and vice versa. In equilibrium, we have that
\begin{equation}
k_{1}n_{1}^{\left(  eq\right)  2}=k_{2}n_{2}^{\left(  eq\right)
}.\label{rates}%
\end{equation}
For the free energy, we use a simple squared-gradient model,
\begin{align}
F\left[  n_{1},n_{2}\right]   &  =\int\left\{  f_{hs}\left(  n_{1}\left(
\mathbf{r}\right)  ;d\right)  +\frac{1}{2}g\left(  \nabla n_{1}\left(
\mathbf{r}\right)  \right)  ^{2}\right\}  d\mathbf{r}\label{Model}\\
&  +\int\left\{  f_{LJ}\left(  n_{2}\left(  \mathbf{r}\right)  \right)
+\frac{1}{2}g\left(  \nabla n_{2}\left(  \mathbf{r}\right)  \right)
^{2}\right\}  d\mathbf{r}\nonumber\\
&  +\int\left\{  f_{hs}^{(ex)}\left(  n_{1}\left(  \mathbf{r}\right)
+n_{2}\left(  \mathbf{r}\right)  ;d\right)  -f_{hs}^{(ex)}\left(  n_{1}\left(
\mathbf{r}\right)  ;d\right)  -f_{hs}^{(ex)}\left(  n_{2}\left(
\mathbf{r}\right)  ;d\right)  \right\}  d\mathbf{r,}\nonumber
\end{align}
As indicated by the notation, the monomers are treated as hard spheres of
diameter $d$ with bulk free energy $f_{hs}(n;d)$ and the dimers as
Lennard-Jones with bulk free energy $f_{LJ}\left(  n;T\right)  $ where $T$ is
the temperature: for the former we use Carnhan-Starling and for the latter a
parameterized form from the literature\cite{JZG}. The squared gradient terms are, as
indicated, taken to have the same coefficient (an assumption that is
artificial but as far as we know, harmless) and this is calculated from the
Lennard-Jones potential\cite{Lutsko2011a}. The final term in the free energy
functional represents the interaction between the two species and for this we
assume that the two species see one-another as hard spheres with the same
diameter as the monomer. Only the excess part of the free energy (i.e.
$f_{ex}\left(  n\right)  \equiv f\left(  n\right)  -f_{id}\left(  n\right)  $,
$\beta f_{id}\left(  n\right)  =n\ln n\Lambda^{3}-n$ where $\Lambda$ is the
thermal wavelength and $\beta=1/k_{B}T$) is used since the ideal parts are
already accounted for. Note that if either density goes to zero, the
interaction term vanishes, as one would expect. The dynamics is given by a
simple generalization of the usual DDFT\cite{MarconiTarazona,EvansArcher,lutsko:acp} to include the reactions,%
\begin{align}
\frac{\partial n_{1}\left(  \mathbf{r;t}\right)  }{\partial t} &
=D_{1}\mathbf{\nabla\cdot}n_{1}\left(  \mathbf{r;t}\right)  \mathbf{\nabla
}\frac{\delta F\left[  n_{1},n_{2}\right]  }{\delta n_{1}\left(
\mathbf{r;t}\right)  }-2k_{1}n_{1}^{2}\left(  \mathbf{r;t}\right)
+2k_{2}n_{2}\left(  \mathbf{r;t}\right)  \\
\frac{\partial n_{2}\left(  \mathbf{r;t}\right)  }{\partial t} &
=D_{2}\mathbf{\nabla\cdot}n_{2}\left(  \mathbf{r;t}\right)  \mathbf{\nabla
}\frac{\delta F\left[  n_{1},n_{2}\right]  }{\delta n_{2}\left(
\mathbf{r;t}\right)  }+k_{1}n_{1}^{2}\left(  \mathbf{r;t}\right)  -k_{2}%
n_{2}\left(  \mathbf{r;t}\right)  .\nonumber
\end{align}
where $D_{1}$ and $D_{2}$ are the tracer diffusion constants for the two
species. At low densities, the free energy can be approximated by the ideal
gas limit,
\begin{equation}
F\left[  n_{1},n_{2}\right]  \simeq\sum_{a=1,2}F_{id}\left[  n_{a}\right]
=k_{B}T\sum_{a=1,2}\int\left(  n_{a}\left(  \mathbf{r}\right)  \ln
n_{a}\left(  \mathbf{r}\right)  \Lambda_{a}^{D}-n_{a}\left(  \mathbf{r}%
\right)  \right)  d\mathbf{r}%
\end{equation}
and the DDFT equations become%
\begin{align}
\frac{\partial n_{1}\left(  \mathbf{r;t}\right)  }{\partial t} &
=D_{1}\mathbf{\nabla}^{2}n_{1}\left(  \mathbf{r;t}\right)  -2k_{1}n_{1}%
^{2}\left(  \mathbf{r;t}\right)  +2k_{2}n_{2}\left(  \mathbf{r;t}\right)  \\
\frac{\partial n_{2}\left(  \mathbf{r;t}\right)  }{\partial t} &
=D_{2}\mathbf{\nabla}^{2}n_{2}\left(  \mathbf{r;t}\right)  +k_{1}n_{1}%
^{2}\left(  \mathbf{r;t}\right)  -k_{2}n_{2}\left(  \mathbf{r;t}\right)
.\nonumber
\end{align}
which are the well-known reaction-diffusion equations. The DDFT model is
therefore a natural generalization of this limit.

We further simplify by taking $D_{1}=D_{2}$. In our previous work, we solved
this model numerically under the assumption of spherical symmetry. In all
cases, the value of the dimer density in the low-density phase is simply taken
from the Lennard-Jones phase diagram to be that of a vapor with a given
supersaturation (we generally worked at supersaturation twice the density of
the vapor in coexistence). We fix the monomer density in the dilute phase by
specifying that it is some multiple of the dimer density (we took it to be
five times the dimer density). Then, the condition for equilibrium,
Eq.(\ref{rates}), means that only one of the rate constants is independent.
The time scale is chosen so that $D_{1}=1$ and lengths are scaled to the
hard-sphere radius. To minimize length scales, we take the latter to be equal
to the length scale of the Lennard-Jones potential. Although these simplifying
assumptions are somewhat artificial, we do not believe they materially affect
the results. Details of the numerical methods can be found in Ref.\cite{LN2}.

Our main conclusion was that this model leads to stable dimer droplets when
the dense dimer phase is stable with respect to the low-density dimer phase.
To understand this, we begin by noting that in the absence of monomer (
$n_{1}=0$), super-critical droplets of the dimer phase grow according
to\cite{Lutsko_JCP_2012_1}
\begin{equation}
\frac{dR}{dt}=aR^{-1},a=Dn_{2}^{\left(  \infty\right)  }\frac{\beta P\left(
n_{2}^{\left(  0\right)  }\right)  -\beta P\left(  n_{2}^{\left(
\infty\right)  }\right)  }{\left(  n_{2}^{\left(  0\right)  }-n_{2}^{\left(
\infty\right)  }\right)  ^{2}} \label{R}%
\end{equation}
where $n_{2}^{\left(  \infty\right)  }$ is the density of the dimer far from
the droplet (i.e. the density of the initial phase), $n_{2}^{\left(  0\right)
}$ is the density of the condensed phase and $P\left(  n\right)
=\frac{\partial f_{LJ}\left(  n\right)  }{\partial n}-f_{LJ}\left(  n\right)
$ is the pressure, giving the classical growth law $R\sim t^{1/2}%
$\cite{Lifshitz}$.$On the other hand, the dimer-rich droplet is out of
equilibrium with respect to the reaction with the monomers:\ dimers will be
converted into monomers in an attempt to reach equilibrium. However, the
monomers so produced will see themselves as hard spheres in a dense
environment and will tend to quickly diffuse out of the droplet. The net
result is that the number of dimers in the droplet is reduced. From the given
reaction, it is easy to show that this leads to a reduction of the radius that
goes as $\frac{dR}{dt}=-k_{2}R/3$. Assuming that both mechanisms operate
additively, the radius obeys $\frac{dR}{dt}=aR^{-1}-k_{2}R/3$ leading to
stabilization at $R=\sqrt{3a/k_{2}}$. These simple predictions are in
reasonable agreement with numerical solution of the DDFT model.

\section{State-dependent reactions}

As stated in the Introduction, one questionable element of the model presented
in the last Section is that the reaction rates are taken to be constant. Here,
we generalize to non-ideal, state-dependent reaction rates. If we assume, as
indicated above, that the reaction rates should depend on the free energy of
the molecules participating in the reaction, then we might suppose that the
rates should go as\cite{nonlinear1,nonlinear2}%
\begin{equation}
r\sim\exp\left(  \beta\sum_{a}|\nu_{a}|\mu_{a}\left(  \mathbf{r}\right)  \right)
\end{equation}
where the sum is over the species participating in the reaction, $\nu_{a}$ is the
stochiometric coefficent  of species $a$ and $\mu_{a}\left(  \mathbf{r}\right)  $ is
the (local) free energy per particle of species $a$, e.g. the local chemical
potential for species $a$. The latter we express using the standard
approximation behind DDFT:\ namely, the assumption of local equilibrium and
DFT as
\begin{equation}
\mu_{a}\left(  \mathbf{r}\right)  =\frac{\delta F\left[  n_{1},...\right]
}{\delta n_{a}\left(  \mathbf{r}\right)  }.
\end{equation}
Note that at low densities, in the ideal gas limit,
\begin{equation}
\frac{\delta F\left[  n_{1},...\right]  }{\delta n_{a}\left(  \mathbf{r}%
\right)  }=k_{B}T\ln n_{a}\left(  \mathbf{r}\right)  \Lambda_{a}^{D}%
\end{equation}
and%
\begin{equation}
r\sim%
%TCIMACRO{\tprod \limits_{a}}%
%BeginExpansion
{\textstyle\prod\limits_{a}}
%EndExpansion
n_{a}\left(  \mathbf{r}\right)  ^{|\nu_{a}|}%
\end{equation}
as expected. Thus, this provides a natural generalization of the usual law of
mass action. Our DDFT model then becomes%
\begin{align}
\frac{\partial n_{1}\left(  \mathbf{r;t}\right)  }{\partial t} &
=D_{1}\mathbf{\nabla\cdot}n_{1}\left(  \mathbf{r;t}\right)  \mathbf{\nabla
}\frac{\delta F\left[  n_{1},n_{2}\right]  }{\delta n_{1}\left(
\mathbf{r;t}\right)  }-2k_{1}e^{2\beta\mu_{1}^{\left(  ex\right)  }\left(
\mathbf{r;t}\right)  }n_{1}^{2}\left(  \mathbf{r;t}\right)  +2k_{2}e^{\beta
\mu_{2}^{\left(  ex\right)  }\left(  \mathbf{r;t}\right)  }n_{2}\left(
\mathbf{r;t}\right)  \\
\frac{\partial n_{2}\left(  \mathbf{r;t}\right)  }{\partial t} &
=D_{2}\mathbf{\nabla\cdot}n_{2}\left(  \mathbf{r;t}\right)  \mathbf{\nabla
}\frac{\delta F\left[  n_{1},n_{2}\right]  }{\delta n_{2}\left(
\mathbf{r;t}\right)  }+k_{1}e^{2\beta\mu_{1}^{\left(  ex\right)  }\left(
\mathbf{r;t}\right)  }n_{1}^{2}\left(  \mathbf{r;t}\right)  -k_{2}e^{\beta
\mu_{2}^{\left(  ex\right)  }\left(  \mathbf{r;t}\right)  }n_{2}\left(
\mathbf{r;t}\right)  .\nonumber
\end{align}
where $\mu_{a}^{\left(  ex\right)  }$ is calculated from the excess part of
the free energy. Written like this, the non-ideal contributions can be viewed
as making the reaction constants, $k_{a}$, state dependent. It is important to note that the generalization used here introduces no new parmameters into the problem: as in our previous work with ideal reactions, the independent parameters are the concentrations (or densities) of the two species in the (meta-)equilibrium dilute phase and, say, the monomer to dimer reacation rate $k_{1}$. The dimer to monomer rate $k_{2}$ is then fixed by the condition for chemical equilibrium between the dilute phases and the diffusion constants are taken to be eliminated by rescaling the time. 

\begin{figure}
[ptb]\includegraphics[angle=0,scale=0.35]{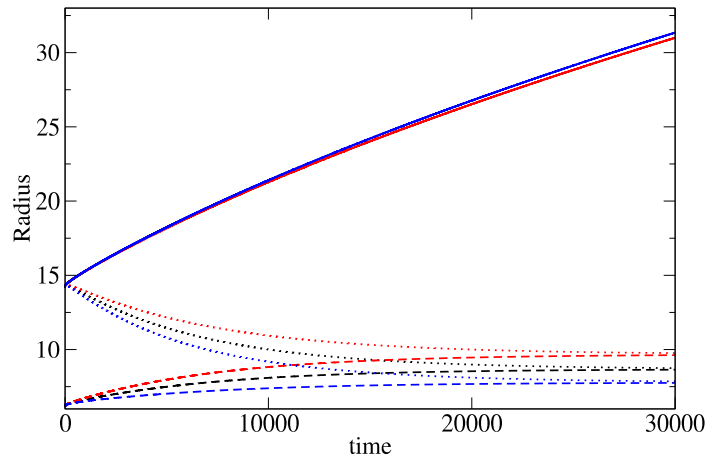}
\caption{Behavior of the cluster radius as a function of time (both in dimensionless units) for three different values of the reaction parameter, $k_{1}^{*} = 8.75 \times 10^{-4}$ (upper curve), $7.5 \times 10^{-4}$ (middle curve) and $10^{-3}$ (lower curve). The broken curves are for the results using constant, state-independent, reaction rates for which  two initial configurations are used: one with a small initial displacement of the critical cluster, and one with a large initial displacement. In all three cases, both initial conditions lead to the same final cluster radius thus demonstrating the stability of the final cluster. The full curves are the results with state-dependent rates and show little sensitivity to the reaction rate and no evident convergence.}
\label{fig_convergence}
\end{figure}

Figure 1 shows numerical results for both models - with constant reaction
rates and with state-dependent rates - using the following protocol. We first
determine the critical cluster for a pure dimer system at the chosen
temperature and background density. We then increase the radius of the cluster
by a specified amount, $\Delta R$. Next, we introduce a uniform background of
monomers. Finally, we evolve this initial system by numerically solving the
DDFT equations. In our previous work, we used two values of $\Delta R$: a
"small" value and a "large" value. We showed that in both cases, the system
evolved toward the same final state which is a cluster larger than the initial
condition for the small value of $\Delta R$ and smaller than the initial
condition for the large value of $\Delta R$. These results are reproduced in
Fig. 1which shows that for three different values of the reaction rate,
$k_{1}$, we obtain final stable clusters with three different radii. Further
analysis shows that the radii are predicted reasonably well by the simple
capillary result, Eq.(\ref{R}).  We also show what happens when we evolve from
the large-$\Delta R$ initial condition using the state-dependent rates. In
this case, there is no evident convergence. In fact, the three clusters grow
at almost identical rates indicating little sensitivity to the reaction rate. Further calculations beginning with much large initial radii, over one hundred length units, give similar results with no sign of stabilization.

\begin{figure}
[ptb]\includegraphics[angle=0,scale=0.35]{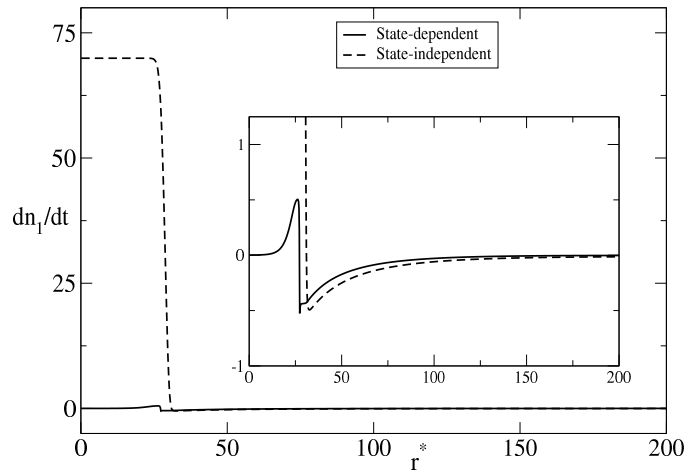}
\caption{Snapshot of the monomer production rate as a function of distance, $r^{*} = r/\sigma$ where $\sigma$ is the Lennard-Jones length scale,  from the center of the cluster for $k_{1}^{*} = 7.5 \time 10^{-4}$. The solid curves show the rate, in dimensionless units, for the case of state-dependent reaction coefficients and the broken curve shows the rate for constant reaction coefficients. }
\label{k_state_2}
\end{figure}

The fact that the state-dependent reaction rates seem to destabilize the
cluster can be understood by estimating their effect in the capillary model.
There, the dimer density inside the cluster is assumed to be that of the
condensed phase under the imposed thermodynamic conditions. If we are at
conditions near coexistence of the dimer vapor and liquid phases, then if the
vapor phase has sufficiently low density, it can be treated as an ideal gas
and we will have that $\beta\mu_{2}^{\left(  \infty\right)  }\simeq\ln
n_{2}^{\left(  \infty\right)  }\Lambda_{2}^{3}$. and for the condensed phase,
$\beta\mu_{2}^{\left(  0\right)  }=\ln n_{2}^{\left(  0\right)  }\Lambda
_{2}^{3}+\beta\mu_{2}^{\left(  0,ex\right)  }$. Assuming, as is true at
coexistence, that these are more or less equal gives $\ln\beta\mu_{2}^{\left(
0,ex\right)  }\simeq\ln\frac{n_{2}^{\left(  \infty\right)  }}{n_{2}^{\left(
0\right)  }}$ so that the reaction term for the dimers inside the cluster will
be
\[
k_{2}e^{\beta\mu_{2}^{\left(  0,ex\right)  }\left(  \mathbf{r;t}\right)
}n_{2}^{\left(  0\right)  }\simeq k_{2}n_{2}^{\left(  \infty\right)  }%
\]
In other words, there is no enhanced conversion of dimers to monomers inside
the cluster, even though the dimer density is much higher than in the
surrounding, low-density background. As a consequence, the process which
balanced the growth of the cluster in the case of constant reaction rates -
the conversion of dimers to monomers that were then expelled from the cluster
for entropic reasons - is eliminated. As a result, the cluster simply grows
like a super-critical cluster. These heuristic estimates are confirmed by detailed examination of the reaction rates as a function of distance in a cluster as shown in Fig.(\ref{k_state_2}). With the constant reaction rates, there is a large excess production of monomers within the cluster but when the state-dependent rates are used, the rate of monomer production within the cluster is two orders of magnitude smaller and, in fact, is largely balanced by a negative production rate (i.e. excess production of dimers) in the interfacial region. 

\section{Conclusion}

In this paper, we have extended our DDFT model for a reacting mixture to
include state-dependent reaction rates. This was motivated by the physical
intuition that reactions should be suppressed when one of the constituents is
in an energetically favorable state in analogy to the way that the rate of
crossing an energy barrier depends on the energy of the initial state.
Numerical calculations based on this model suggest that such a modification
eliminates the mechanism that was previously identified - for the case of
constant reaction rates - as stabilizing a post-critical cluster against
continued growth. Finally, we gave a simple argument that indicates why this
happens: the dependence of the rates on the local chemical potential means
that the over-all rate of conversion of dimers to monomers is insensitive to
the dimer density so that there is no preferential conversion within a
cluster. This argument relies on conditions being close to coexistence for the
two dimer phases, so it may leave open the possibility of stabilization for
some particular conditions. However, at the very least, it seriously
compromises the robustness of the mechanism. 

Within the context of the capillary model, it might seem that the
state-dependent rates only change the size of the cluster since the mechanism
opposing growth is the conversion of dimers to monomers which are subsequently
expelled by diffusion. This leads to a net rate of loss of dimers satisfying
$dN_{2}/dt=-k_{2}N_{2},$where $N_{2}$ is the total number of dimers in the
cluster,  which is then balanced by the tendency of the supercritical cluster
to grow. Based on the arguments given above, the state dependent rates would
seem to modify this to give $dN_{2}/dt=-k_{2}\left(  \frac{n_{2}^{\left(
\infty\right)  }}{n_{2}^{\left(  0\right)  }}\right)  N_{2}$ which simply
amounts to a change in the effective reaction rate. While this is true, one
must also consider the effect on the monomers. Their reaction rate $k_{1}$ is
increased due to the fact that there is a high free energy cost to insert a
monomer into the cluster. As a consequence, they are rapidly converted back
into dimers so that the net change in the number of dimers due to the reaction
is small. 

Of course, the original conclusion that the mechanism is valid could still hold if the introduction of state-dependent reaction rates is a
choice:\ if one could imagine a system in which, somehow, such a dependence
is not present. However, we believe this is unlikely. The state-dependent
rates are a realization of detailed balance which is generally necessary to
allow for a relaxation to an equilibrium state. In the present problem, this
manifests itself in a more direct manner since it addresses a fundamental
physical weakness of the proposed stability mechanism. Although our
realization here differs in some details from the original proposal of Vekilov
et al, it retains the fundamental feature that in the case that the protein
clusters are stabilized, there is a steady-state flow between the cluster and
the background solution. In our case, with constant reaction rates, this is a
result of the process by which the excess of dimers in the cluster lead to the
rapid conversion of dimers to monomers which are then expelled by diffusion
(since the monomers behave as hard spheres in a dense environment). This
causes the cluster to become smaller but is balanced by the fact that the
super-critical dimer cluster is driven to grow so as to reduce its free
energy. The latter process involves the absorption of dimers by the cluster
thus setting up a closed loop: dimers enter the cluster, are converted to
monomers and finally  are expelled. This represents a nonequilibrium steady
state that should not be sustainable without a source of energy as was noted
in Ref.(\cite{LN2}) where it was speculated that for this reason, the
clusters might only be stable on certain time scales. However, based on the
present work, it seems more likely that it was the violation of detailed
balance that led to the artificial stability of the nonequilibrium steady
state and that once this unphysical feature is corrected, there can be no
self-sustaining currents and, hence, no cluster stability. 

Our results are negative: we show that while the proposed theoretical model based on a conversion between proteins and some type of complex is viable for state-independent reaction rates, it seems to fail when non-ideal reactions are considered. Negative results such as these cannot be definitive: it is always possible that there is some set of paramter values or some change to the assumed interaction potentials that will restore stability. However, the fact that the effect of non-ideal intereactions is so dramatic - reducing the overall conversion rate from dimers to polymers within the cluster by a factor of more than 100 - suggests that no small change is likely to restore stability. The theoretical arguments given above support the notion that the dimer to monomer rate will be reduced to a value similar to that in the dilute solution outside the cluster while the monomer to dimer rate will be much higher leading to a very low overall monomer to dimer rate regardless of the paramters such as the temperature or the densities in the dilute phases.  We also note that one of the virtues of the proposed model with \emph{ideal} reaction rates is that it is very robust: our previous investigation of the model showed that it was relatively insensitive to details of the model. Since protein clusters are observed under a wide range of conditions and for various proteins, it seems unlikely that the mechanism can require sensitive tuning of its parameters.

In conclusion, Pan et al suggested a theoretical model to explain the stablility of protein clusters as observed in experiment. The model presupposes  the presence of an additional species that is in equilibrium with the protein monomers and that condenses. Cluster stability is obtained through a balance of supercritical growth and conversion from the new species back to monomers within the cluster. The results presented here clearly challenge this model and suggest that it cannot explain the stability of the clusters observed in experiment. We argue that the original mechanism relied on an unphysical steady, nonequilibrium process that should not be possible without an external driving force and that this was in fact a manifestation of the fact that ideal interactions violate detailed balance. When detail balance is restored via state-dependent interactions, the nonequilibrium state no longer exists and stability is lost. The unquestioned stability observed in experiments could be the result of the present model with additional physics such as multiple oligimers with different interaction potentials (see e.g. Ref.\cite{additional}) or to some completely new mechanism. The theoretical explanation for the observations therefore remains an open question.

\begin{acknowledgments}
This work was supported in part by the European Space Agency under contract
number ESA AO-2004-070. The author thanks Pierre Gaspard for suggesting the importance of non-ideal affinities and Gregoire Nicolis for on-going discussions. The author also thank Peter Vekilov for numerous
conversations on this subject. 
\end{acknowledgments}

\bibliography{protein_clusters}

\end{document}